\documentclass[preprint]{nature}%
\usepackage{graphicx}
\usepackage[usenames,dvipsnames]{color}
\usepackage{siunitx}
\usepackage[colorlinks=true,citecolor=blue,linkcolor=blue,urlcolor=blue]%
{hyperref}
\usepackage[a4paper]{geometry}
\usepackage{fixltx2e}
\usepackage{upgreek,xspace}
\usepackage{amsmath}
\usepackage{amssymb}
\usepackage{amsthm}
\usepackage{mathrsfs}
\usepackage{amsbsy}
\usepackage{amstext}
\usepackage{xcolor}
\usepackage{float}
\usepackage{multirow}
\usepackage{wasysym}
\usepackage{mathtools}
\usepackage{subfigure}
\usepackage{amsfonts}
\usepackage{dcolumn}
\usepackage{float}
\usepackage{caption}
\usepackage[right]{lineno}
\usepackage{fancyhdr}
\usepackage{bbold,bm}
\usepackage[utf8]{inputenc}
\usepackage[T1]{fontenc}%
\usepackage{makecell}  
\usepackage{threeparttable}
\usepackage{booktabs,caption}
\providecommand{\U}[1]{\protect\rule{.1in}{.1in}}
\DeclareCaptionLabelSeparator{vline}{ $\boldsymbol{\vert}$ }
\usepackage{soul}
\usepackage{physics}

\begin{document}

\title{Solution-phase single-particle spectroscopy for probing multi-polaronic dynamics in quantum emitters at femtosecond resolution}

\maketitle
\author{Jiaojian Shi$^{1,2, \dagger}$, Yuejun Shen$^{1,2, \dagger}$, Feng Pan$^{1, \dagger}$, Weiwei Sun$^{3}$, Anudeep Mangu$^{1,2}$, Cindy Shi$^{1}$, Amy McKeown-Green$^{4}$, Parivash Moradifar$^{1}$, Moungi G. Bawendi$^{3}$, William E. Moerner$^{4}$, Jennifer A. Dionne$^{1,5}$, Fang Liu$^{4}$, Aaron M. Lindenberg$^{1,2,6,*}$}

\begin{affiliations}
	
\item Department of Materials Science and Engineering, Stanford University, Stanford, California 94305, United States
\item Stanford Institute for Materials and Energy Sciences, SLAC National Accelerator Laboratory, Menlo Park, California 94025, United States
\item Department of Chemistry, Massachusetts Institute of Technology, Cambridge, Massachusetts 02139, United States
\item Department of Chemistry, Stanford University, Stanford, California 94305, United States
\item Department of Radiology, Stanford University, Stanford, California 94305, United States
\item Stanford PULSE Institute, SLAC National Accelerator Laboratory, Menlo Park, California 94025, United States
\\
	$^{\dagger}$These authors contributed equally\\
	$^{*}$Email: A.M.L. (aaronl@stanford.edu)
\newpage
	\vspace{1cm}
\end{affiliations}
\normalsize
\begin{abstract}
The development of many optical quantum technologies depends on the availability of solid-state single quantum emitters with near-perfect optical coherence. However, a standing issue that limits systematic improvement is the significant sample heterogeneity and lack of mechanistic understanding of microscopic energy flow at the single emitter level and ultrafast timescales. Here we develop solution-phase single-particle pump-probe spectroscopy with photon correlation detection that captures sample-averaged dynamics in single molecules and/or defect states with unprecedented clarity at femtosecond resolution. We apply this technique to single quantum emitters in two-dimensional hexagonal boron nitride, which suffers from significant heterogeneity and low quantum efficiency. From millisecond to nanosecond timescales, the translation diffusion, metastable-state-related bunching shoulders, rotational dynamics, and antibunching features are disentangled by their distinct photon-correlation timescales, which collectively quantify the normalized two-photon emission quantum yield. Leveraging its femtosecond resolution, spectral selectivity and ultralow noise (two orders of magnitude improvement over solid-state methods), we visualize electron-phonon coupling in the time domain at the single defect level, and discover the acceleration of polaronic formation driven by multi-electron excitation. Corroborated with results from a theoretical polaron model, we show how this translates to sample-averaged photon fidelity characterization of cascaded emission efficiency and optical decoherence time. Our work provides a framework for ultrafast spectroscopy in single emitters, molecules, or defects prone to photoluminescence intermittency and heterogeneity, opening new avenues of extreme-scale characterization and synthetic improvements for quantum information applications.
\end{abstract}

\newpage


A frontier of quantum information science research is the development of non-classical light sources that produce streams of photons with controllable quantum correlations. A central building block is scalable single quantum emitters (SQEs) capable of producing indistinguishable single photons or entangled photon pairs on demand~\cite{Brien2009,Lodahl2015,Aharonovich2016,Senellart2017}. Such SQEs play a significant role in a range of proposed quantum computing, cryptography, and communication applications~\cite{Aspuru-Guzik2012,Scarani2009,Gisin2007} and have been explored in various material systems~\cite{Castelletto2014,Michler2000,Ma2015, He2015,Koperski2015,Chakraborty2015,Srivastava2015}. Notable examples include emissive defects in two-dimensional hexagonal boron nitride (hBN) that exhibit promising quantum-optical properties at room temperature, including high photon-extraction efficiency, narrow linewidth, and easy integrability~\cite{Tran2016,Grosso2017}.

However, the currently available hBN SQEs suffer from low quantum efficiency and rapid decoherence~\cite{Spokoyny2020,Xu2021}. The main obstacles to their synthetic improvement are the significant sample heterogeneity and lack of fundamental understanding of the nanoscale energy flow that determines their novel macroscopic functionalities, such as optical coherence. Considerable variations between hBN SQEs have been reported in their emission spectrum~\cite{Tran2016c,Chejanovsky2016,Hayee2020}, quantum yield (QY)~\cite{Xu2021,Ziegler2019,Li2021}, photostability~\cite{Shotan2016,Exarhos2017}, susceptibility to excitation wavelengths~\cite{Schell2018}, etc. Such heterogeneity is also omnipresent in a wide range of physical quantities of many other single emitters at nearly all timescales. For instance,  (1) The decoherence caused by microsecond-to-millisecond ($\mu$s-ms) photoluminescence (PL) blinking is universal~\cite{Frantsuzov2008,Shi2021} but exhibits disparate behaviors between SQEs~\cite{Li2021,Ziegler2019}; (2) The two-photon (TP) emission probability governs the single-photon purity and entanglement of quantum emission cascade~\cite{Senellart2017,Oliver2000}. However, its dynamics at nanosecond (ns) to $\mu$s timescales are highly susceptible to defects and dependent on the local environment, making it challenging to measure~\cite{Nair2011,Zhao2012,Shotan2016,Feldman2019}; (3) Microscopic views of excited-state relaxation and electron-phonon coupling are keys to understanding quantum light emission and its decoherence mechanism~\cite{Pan2012,Senellart2017}, but their dynamics at the femto- to picosecond timescales are found to vary dramatically between different single emitters or molecules~\cite{Moya2022,Van2005,Grosso2020}.


Therefore, it is imperative to develop single-particle spectroscopy that can interrogate over many SQEs with temporal resolution at the femtosecond regime. One strategy is to build on advances in solution-phase fluorescence correlation (S-g$^{(2)}$) analysis and single-molecule pump-probe (SMPP) spectroscopy. The S-g$^{(2)}$ method has been adopted in revealing ensemble-level ns-ms dynamics in colloidal nanocrystal research~\cite{Beyler2014,Cui2014}. Meanwhile, SMPP spectroscopy can capture ultrafast femtosecond-to-picosecond (fs-ps) dynamics in a single molecule or emitter but suffers from significant sample heterogeneity, photobleaching, and consequent poor signal-to-noise, especially at room temperature~\cite{Moya2022,Van2005,Maly2016,Hildner2013,Brinks2010,Preuss2022}. The marriage of these two techniques in SQE research could provide an unprecedented dynamical view of sample-averaged physical quantities spanning up to 15 orders of magnitude in time. The immediate and reliable feedback from this method also has the potential to provide facile scrutiny of microscopic energy flows and synthetic guidance for improved emission quality, but has never been demonstrated. 



Here we develop solution-phase single-molecule pump-probe (S-SMPP) spectroscopy and probe the sample-averaged photon-correlation dynamics in hBN SQEs down to fs resolution. The particle Brownian diffusion through the laser focal volume allows sample-averaged characterization over an entire batch of samples. By analyzing photon antibunching at ns timescales and sub-ms bunching mainly from emitter diffusion through the laser focus, the sample-averaged normalized TP QY of hBN SQEs can be determined to have an upper limit of 12.8$\pm$2$\%$. Two bunching shoulders at about tens of ns and sub-$\mu$s are highly dependent on polarization configurations and fluences, indicating their origins from SQE rotational dynamics and metastable shelving states, respectively. We also show that S-SMPP with fluorescence correlation detection can isolate genuine SQE emission from uncorrelated background counts and capture multi-polaronic dynamics in hBN SQEs with ultrafast resolution ($\sim$ 100 fs) and ultrawide temporal spanning (simultaneously covering 15 orders of magnitude in time). The further discovery of accelerated polaronic formation upon multi-electron excitation is supported by theory and can translate to evaluating the force exerted by an electron on the lattice and the optical coherence time. Such a solution-phase approach also enables short exposure times to each SQE and alleviates experimental concerns of emission intermittency and bleaching, which allows a time-dependent view of enhanced lattice deformation under an intense multi-electron excitation regime. Compared with conventional single-color SMPP, this method exhibits advantages in removing coherent artifacts and achieving an ultralow noise level and can readily translate to study other single emitters or molecules prone to PL blinking/bleaching or with physicochemical properties modified by coating on a substrate or non-native environment~\cite{Coleman2011,Shirasaki2013,Galland2011}.

The S-g$^{(2)}$ setup schematic is shown in Fig.~1a. The sample consists of hBN nanoflakes in solution which are excited by a continuous-wave (CW) or pulsed 532-nm laser beam with PL collected by a confocal microscope and detected by a pair of avalanche photodiodes in a Hanbury Brown and Twiss interferometer configuration. Figure~1b shows a typical PL trace as a function of experimental time. The spikes in PL counts are mainly due to the nanoflakes diffusing through the laser focal volume via Brownian motion. It may also contain a contribution from PL blinking, which occurs at similar ms timescales~\cite{Shotan2016,Tran2016,Kianinia2018}. Figure~1c shows the S-$g^{(2)}$ trace at the timescale of 10$^{-6}$-10$^{-1}$ s, which exhibit a gradual bunching feature at $\tau$ < 1 ms and convergence to g$^{(2)}$ = 1 at $\tau$ > 100 ms. This trace can be fitted with a 3D normal diffusion equation used in fluorescence correlation spectroscopy (FCS)~\cite{Krichevsky2002,Bacia2006} (details in Supplementary Note~2), which yields a diffusion time $\tau_D = 1.6\pm0.09$ ms and average occupation of particles in the focal volume $\bar{N} = 0.947\pm0.003$. Such correlation analysis can be understood as an active normalization of coincidence counts with samples (SQE in the focal volume) to the reference coincidence counts (SQE outside the focal volume) enabled by particle diffusion motions. Therefore it allows the Poisson background contributions, either environmental scattering or solution autofluorescence, to be canceled, and serves as the basis for sample-averaged characterization in solution. 

Figure~2a shows the S-g$^{(2)}$ trace covering a time window from 10$^{-10}$ to 10$^{-1}$ s. We observe three features beyond the classical diffusion dynamics (marked as region I) --- antibunching at $\tau$ < 10$^{-8}$ s (marked as IV); a correlation spike at  $\tau\sim2\times$10$^{-8}$ s (marked as III); a bunching shoulder at $\tau\sim2\times$10$^{-8}$ -- $1\times$10$^{-6}$ s (marked as II). A detailed view of region IV and its single exponential fit is shown in Fig.~2b. The decay lifetime is $2.8\pm0.3$ ns, which agrees with the PL lifetime ($\sim$ 3 ns) reported in other literature~\cite{Tran2016}. The antibunching of hBN SQEs in solution does not approach zero as it would in a single-molecule g$^{(2)}$ experiment on a substrate because there is a Poisson distribution of particles in the focal volume with a uniform probability at all $\tau$ of measuring photon pairs produced by different particles. This probability, given by the long $\tau$ limit of g$^{(2)}(\tau)$ is normalized to unity instead of zero in FCS. This unity Poisson background from the inter-particle contribution to g$^{(2)}(\tau)$ can be subtracted to isolate the single-particle contribution and estimate the ratio between TP QY and single-photon (SP) QY, as reported in prior literature~\cite{Beyler2014}, $\frac{<\gamma_{SP}\gamma_{TP}>}{<\gamma_{SP}^2>} = \frac{g^{(2)}(\tau = 0) - 1}{g^{(2)}(T) - 1}$, where $\gamma_{SP}$ is SP QY, $\gamma_{TP}$ is TP QY, $T$ is given by the correlation time much larger than PL lifetime ($T\gg3$ ns) and much shorter than the particle diffusion time ($T\ll10$ $\mu$s), and typically chosen to be $T = 100$ ns. Thus the ratio between the $A$ and $B$ values in Fig.~2b characterizes the normalized TP QY: QY(TP)/QY(SP)$ = 12.8\pm2$\%. This value is robust to excitation fluences with no apparent dependence when varied by more than an order of magnitude (see Fig.~S8). Since we cannot exclude the possibility of each nanoflake hosting more than one SQE, this value is an upper limit of the normalized TP QY.

Besides the antibunching feature, a bunching shoulder followed by a spike in the intensity correlation is observed at regions II and III, respectively. The correlation spike shows up at the timescale of tens of ns, where often the rotational dynamics reside. Further polarization-dependent experiments monitor the $g^{(2)}(\tau)$ under varied incident and detected light polarization configuration of $IJK = xyx$, where $IJK$ represents the correlation of the detected $J$-polarized signal with the detected $K$-polarized signal under excitation polarization $I$. Figure~2c shows that the spike almost disappears with a polarization configuration of $IJK = xyx$ and provides additional evidence of its rotational origins. As previous literature on other molecular systems reported, they are attributed to the negative rotational correlation between two channels with orthogonal polarization states~\cite{Kask1989,Oura2016,Dorfschmid2010}. With increasing excitation fluence, Fig.~2d further shows that the rotational correlation spike is strongly suppressed, and the bunching shoulder (hump at region II) is much more prominent. Similar power-dependent bunching shoulders have been observed in some individual hBN SQEs on a substrate and attributed to the metastable (up to 1 $\mu$s lifetime) shelving states~\cite{Shotan2016,Koperski2021,John2017,Neu2011}. The concurrent spike suppression and hump enhancement can be better visualized in Fig.~2e by normalizing the correlation counts in the spike and hump to those in the FCS curve. Such behavior is because the rotation-induced correlation spike relies on the absorption and emission transition dipole moment at specific and well-defined directions~\cite{Kask1989}. They are randomized at longer $\mu$s timescales by the relaxation pathways mediated by diverse shelving states~\cite{Steiner2015}. Figure~2f-g shows S-g$^{(2)}$ measurements at different excitation wavelengths. When illuminated by shorter-wavelength light at 450 nm and 400 nm, the $B$ value is 3 and 10 times lower than that under 532-nm pumping, and average PL counts are considerably lower than that with 532-nm excitation at similar pump fluence (see Fig.~S3). Such dramatic wavelength dependence in SP QY has been similarly observed in individual hBN SQEs, likely due to a more substantial photobleaching effect excited with ultraviolet light~\cite{Exarhos2017, Shotan2016}.

The ns-ms dynamics obtained paved the way to develop tailored S-SMPP spectroscopy that can resolve intricate and rapid energy redistribution between electron and lattice subsystems in hBN SQEs at femtosecond resolution. The setup schematic in Fig.~3a shows the double-pulse excitation scheme incorporated with photon correlation detection. By scanning the pump-probe delays, SMPP can record the transient absorption change encoded in Stokes-shifted fluorescence, directly reporting the dynamical evolution of the excited-state potential surface (inset of Fig.~3a) or energy transfer~\cite{Van2005,Moya2022} within the probe’s spectral window. Its combination with the solution-phase method provides ultrafast characterization over millions of highly heterogeneous hBN SQEs without concerns of PL blinking or photobleaching. The PL detection in S-SMPP poses significant advantages over the absorption observable here, which would be dominated by the host material or non-emissive defects instead of SQEs. By monitoring the second-order fluorescence correlation, the Brownian diffusion motions give rise to the ms bunching features (shown in Fig.~1c), which can be separated from the uncorrelated fluorescence background to represent the pure SQE emission intensity.


Figure~3b shows the emission spectrum from the hBN SQE solution measured with a grating-based spectrometer. From this intensity-based spectral characterization, the PL from hBN SQEs is buried in the solvent Raman scattering signal (detailed assignments are shown in Fig.~S10) and background light. To unravel the genuine SQE emission, three spectral regions marked in Fig.~3b were selected for PL correlation characterization. Figure~3c shows that both the bunched coincidence photon counts (above the dashed line) and overall coincidence counts in the region I are appreciably more than those in region II, which means hBN SQEs preferably emit light with energies located in the region I. Almost no bunched coincidence photon counts are observed in region III, concurrent with high overall coincidence counts. Such corroborated scrutinization of the origins of coincident photon pairs unambiguously assigns the dominating photon counts in the region III to Raman scattering. Figure~3d further shows wavelength-dependent total counts and bunched photon counts. The former observable closely resembles the PL spectrum obtained with the conventional spectrometer with dominating Raman scattered intensity at about 630 nm and 650 nm. These peaks disappear using the latter approach with an emergence of a broad feature peaked at 570 nm expected for hBN SQEs~\cite{Spokoyny2020,Hayee2020}, which represents the sample-averaged hBN SQE emission disentangled from those interfering contributions. This hBN peak spans a wide spectral range from 555 – 590 nm due to interrogation over many heterogeneous SQEs. The stark contrast with the reported narrow emission lines~\cite{Tran2016,Spokoyny2020,Shotan2016} highlights the importance of this sample-averaged characterization tool. We further compare S-SMPP with SMPP using CdSe/CdS quantum dots (QDs) as a reference in Supplementary Note~6 and Fig.~S9. The results show that the phase coherence artifacts in measuring a single emitter on a substrate can be safely destroyed in solution-phase measurements, which sample a volume larger than the interference volume.

Figure~4a shows the ultrafast dynamics obtained from S-SMPP measurements on the solution-phase hBN SQEs. A characteristic PL intensity dip emerges near delay zero between 515-nm pump-probe pulse pairs with $\sim$ 12 mJ/cm$^2$ excitation fluence. The noise level is $\sim$ 0.2\% and about two orders of magnitude lower than that from previous single-color SMPP work (10 - 30\%) done on a single molecule on a substrate~\cite{Moya2022,Maly2016,Van2005}, which is limited by the detrimental PL blinking and photobleaching phenomena. The fit function was the convolution of the measured laser pulse duration with an exponential rise function for energy relaxation (details described in Supplementary Note~4)~\cite{Moya2022,Maly2016}. The extracted excited-state decay time constant $t_1$ is about $30\pm14$ fs and found to be further prolonged at lower excitation fluences, as shown in Fig.~4b-c. These timescales exhibit a stark difference when compared to the decay times from S-SMPP measurements on CdSe/CdS QDs (see Supplementary Note~6). The overall PL intensity suppression is further verified to originate from hBN SQEs diffusing across the laser focal volume by the photon correlation analysis (see Fig.~S11), which captures an apparent reduction of bunched photon coincident events when pump and probe pulse overlap in time. By assuming a similar absorption cross-section of hBN SQEs to the nitrogen-vacancy centers in a diamond~\cite{Wee2007}, average carrier density per SQE (<$N$>) can be estimated to be $\sim 1.8, 0.9, 0.4$, corresponding to the excitation fluence of 12, 6.2, 2.9 mJ/cm$^2$, respectively (details included in Supplementary Note~1). Since the defect-mediated hBN SQEs can be considered as a molecular system with vibrational states coupled to electronic degrees of freedom~\cite{Grosso2020,Sio2023}, the emission spectrum highly depends on their interaction with phonon modes, i.e., polaron formation (displaced potential surface is shown in the inset of Fig.~3a). Thus, the evident reduction of $t_1$ shown in the inset of Fig.~4c translates to an acceleration of polaronic formation and structural distortion at higher <$N$>.

Such fluence-dependent phenomena can be explained by multiple excitations, e.g., bipolarons, that lead to stronger interactions with the lattice, causing a more rapid shift of excited-state energies away from the probe-induced stimulated emission~\cite{Fisher1989}, as shown in Fig.~4d (more discussion in Supplementary Note~7). Following a modified Fr\"ohlich polaron model, we formulated this description by invoking a linearly coupled electron-phonon system and solving the time evolution of energy transfer between subsystems~\cite{Brown1986,Brown1987}. A detailed analysis is provided in Supplementary Note~3. Upon photoexcitation, the lattice energy increases and saturates at an asymptotic value with total energy conserved, as shown in Fig.~4e. Since <$N$> is on the order of unity, more excited electrons at the same defect sites were assumed at higher fluences for the subsequent generation of non-negligible multi-polaron populations. The multi-electron excitation reduces the time for polaron energy to relax out of the laser energy bandwidth ($\sim$ 20 meV), as marked by the dashed lines in Fig.~4e. The simulated relaxation times $t_1=110$ fs (69 fs) for <$N$> = 1 (2) qualitatively agree with the experimental data $t_1 = 48$ fs (30 fs) within an order of magnitude. The corroborated experimental results and theoretical modeling allow further estimation of electron-phonon coupling strength $g \sim 50$ piconewtons for <$N$> = 1, i.e., the classical force an electron exerts on the lattice shown in Fig.~4d. They also yield the spatial-dependent strain evolution upon photoexcitation of a specific site, as shown in Fig.~4f. Within the experimental time $\sim$ 60 fs, the local strain amplitude on the initially excited site is $\sim$ 4\% for <$N$> = 1 (see Fig.~S12 for <$N$> = 2). We expect future development of a more advanced model (e.g., with first-principle calculated optical phonon branches, extension to actual quasi-3D structure including various defect states, etc.) could provide a more accurate description of the multi-polaron formation dynamics. We further note that the measured $t_1$ can translate to the electron-phonon coupling coefficient intimately linked to the pure dephasing time ($T_2’$) interrogated over many SQEs (more discussion in Supplementary Note~3)~\cite{Takagahara1999}. Combined with the spontaneous emission lifetime ($T_1$) measured in Fig.~2b, optical coherence time ($T_2$) that governs the core of quantum computation, teleportation, metrology, etc., can be extracted for guiding rational design of hBN SQEs or new source discoveries.

In conclusion, we have demonstrated that combined S-g$^{(2)}$ and S-SMPP with photon correlation detection can probe sample-averaged dynamics in hBN SQEs at room temperature with unprecedented temporal resolution from ms down to fs. The normalized TP QY can be measured over millions of SQEs by unraveling multiple ms-ns dynamics in time. Spectrally-resolved photon bunching analysis yields hBN SQE emission spectrum free from background scattering. At the ps-fs timescale, the polaronic formation dynamics extracted show apparent acceleration at increasing numbers of electronic excitation, which can be described analytically and utilized to evaluate the benchmark optical coherence. Such a solution-phase method shows the particular merits of automatic sample replenishment for sample-averaged bleaching-free characterization in hBN SQEs and cannot be paralleled with measurements on densely packed thin-film nanoflakes. In that case, significant wavefunction overlap between emitters leads to other relaxation channels like inter-emitter charge transfer~\cite{Pein2017,Wood2011}, and dramatic substrate-dependent phenomena have been revealed~\cite{Shirasaki2013,Hans1970,Galland2011}. S-SMPP on hBN SQE with PL observable further distinguishes itself from conventional transient absorption spectroscopies, which are complicated or blinded by the responses from non-emissive defects and host absorption. In contrast to the ultrafast PL spectroscopies based on Kerr or nonlinear gating techniques~\cite{Takeda2000,Hu2016,Peon2001}, such a method can achieve ps-fs resolution without sacrificing photons, which enables photon-counting of the weak luminescence signal from a single molecule or emitter. S-SMPP can be readily upgraded to achieve versatile ultrafast spectroscopies and coherent controls, e.g., excitation at tailored frequency and chirality; state-selective PL detection; spectral-resolved probing scheme; single-photon purity manipulation; etc. We anticipate this tool's reliable performance and ultrafast temporal resolution will assist the ongoing synthetic effort to improve SQEs in low-dimensional systems~\cite{Coleman2011,Nicolosi2013} and solve emergent theoretical puzzles~\cite{Hayee2020,Li2021,Grosso2017,Li2019}. This approach can also translate to other single biological molecules that are air-/photo-sensitive or difficult to prepare in solid form, challenging to study even under state-of-the-art ultrafast techniques~\cite{Hildner2013,Brinks2010,Piatkowski2019,Xu2021b}. For example, providing ultrafast “snapshots” of quantum coherence in photosensitive biological molecules in their native environments for quantum biology applications~\cite{Scholes2017,Moya2022}; high-throughput in-vivo/in-vitro cellular screening of electron/proton-bath interactions at physiological temperatures with biomedical implications~\cite{Lippincott-Schwartz2001,Li2011}.

\begin{addendum}
	\item[Acknowledgements] J.S., F.P., J.A.D., and A.M.L. acknowledge support from the U.S. Department of Energy, Office of Science, National Quantum Information Science Research Centers. Y.S. acknowledges support from the Department of Energy, Office of Science, Basic Energy Sciences, Materials Sciences and Engineering Division, under Contract DE-AC02-76SF00515. W.S. and M.G.B. acknowledge support from the U.S. Department of Energy, Office of Basic Energy Sciences, Division of Materials Sciences and Engineering (Award DE-SC0021650). This material is based upon work supported by the U.S. Department of Energy, Office of Science, National Quantum Information Science Research Centers. The authors also acknowledge helpful discussions with Hendrik Utzat.
 
	\item[Author contributions] $\dagger$ J.S., Y.S. and F.P. contributed equally to this work. J.S. and A.M.L. conceived the study. J.S., Y.S. and F.P. conducted the experiment under the supervision of A.M.L., F.L. and J.A.D. F.P. and J.S. interpreted S-g$^{(2)}$ data under the guidance of W.E.M. Y.S. and J.S. performed the polaronic simulation to the S-SMPP data under the guidance of A.M.L. W.S. provided the CdSe/CdS QDs under the supervision of M.G.B. C.S. prepared diluted QD solution. C.S. and P.M. performed transmission electron microscope characterization. W.S., A.M. and A.M.G. contributed to data interpretation. J.S., Y.S., and F.P. wrote the manuscript with crucial inputs from all authors.
	
	\item[Competing interests] The authors declare no competing financial interests. 
	
	\item[Additional information] Supplementary Information is available for this paper. Correspondence and requests for materials should be addressed to A.M.L. (aaronl@stanford.edu).
	
	\newpage
	\clearpage
	
\begin{center}
	\includegraphics[width=\columnwidth]{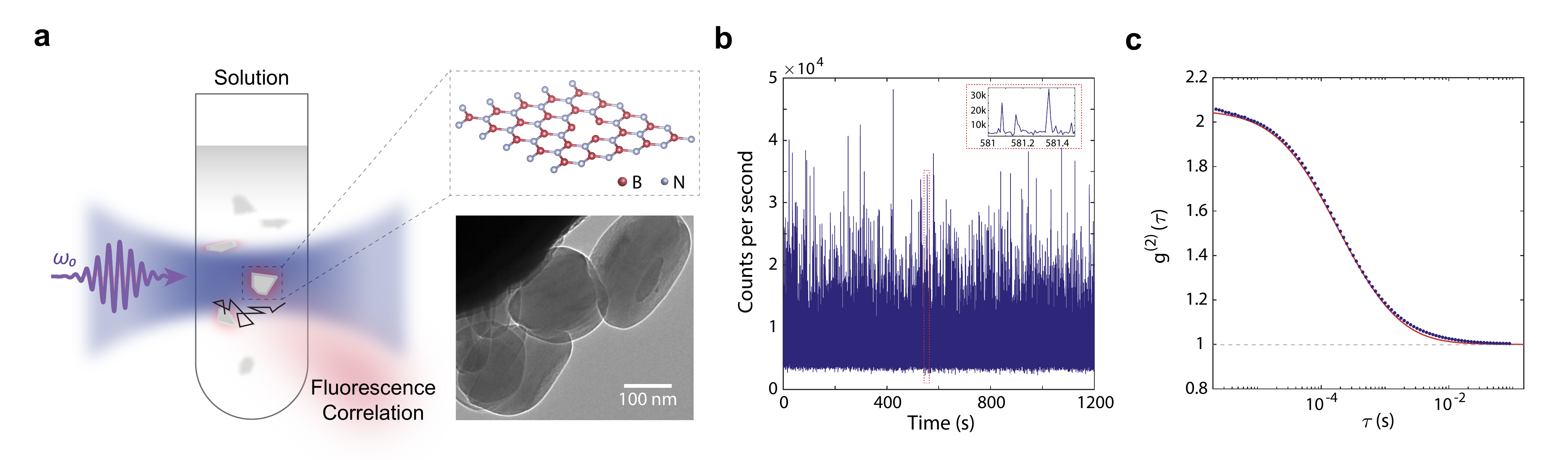}
\end{center}
\item[Fig. 1] \textbf{a}, Schematic illustration of S-$g^{(2)}$ measurement of hBN SQEs in solution. The insets are the lattice structure and transmission electron microscope image of hBN nanoflakes. \textbf{b}, PL counts per second as a function of time show fluctuating intensity due to nanoflakes diffusing through the excitation laser focus. The inset shows a zoom-in PL trace with a window of $\sim$ 0.5 s. \textbf{c}, Normalized intensity correlation function S-$g^{(2)}$ at $\tau$ > 10$^{-5}$ s. The trace can be fit by the translational diffusion dynamics and converges to 1 at $\tau$ > 10$^{-2}$ s.
\newpage
\clearpage

\begin{center}
	\includegraphics[width=\columnwidth]{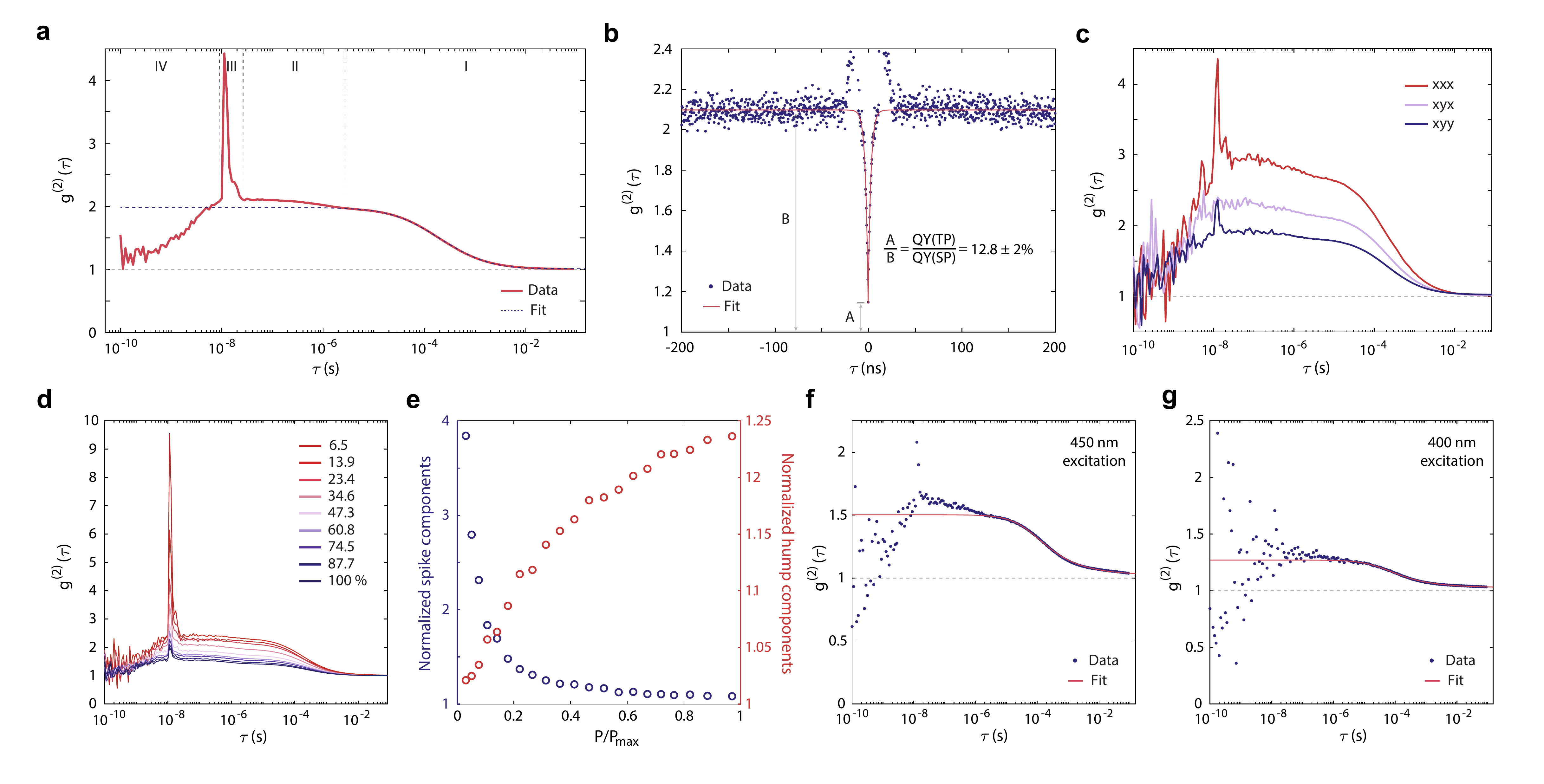}
\end{center}
\item[Fig. 2] \textbf{a}, S-$g^{(2)}$ trace with a window from 10$^{-10}$ to 10$^{-1}$ s. Besides the diffusion dynamics (> 10$^{-5}$ s, marked as region I), two bunching shoulders ($2\times$10$^{-8}$ -- $1\times$10$^{-6}$ s and $\sim2\times$10$^{-8}$ s, marked as region II and III) can be observed and corresponds to metastable shelving states and rotational dynamics, respectively. The antibunching feature at the timescale of < 10$^{-8}$ s (marked as region IV) demonstrates the quantum nature of the emission. \textbf{b}, $g^{(2)}(0)$ above the unity is indicative of TP fluorescence and the $g^{(2)}$($\tau$ > 100 ns) sample the FCS correlation function. \textbf{c}, Polarization-dependent S-$g^{(2)}$ measurement. The excitation polarization is fixed to $x$-polarized. The polarization of the analyzers of the two avalanche photodiodes is set to $xx$, $yx$, and $yy$. The rotation-induced bunching features (spike at $\sim$ 10$^{-8}$ s at region III) are strongly suppressed at the $xyx$ configurations. \textbf{d}, Fluence-dependent S-$g^{(2)}$ measurement shows that the correlation spike is strongly suppressed under increasing excitation powers. \textbf{e}, Progression of normalized time-integrated correlation amplitudes for the spike (blue dot) and hump (red dot). The spike components are normalized to the correlation probability density at later $\tau $ = 10$^{-7}$ s, and the shelving-state-induced components (hump at region II) are normalized to the correlation probability density extracted from translational FCS fits. At increasing fluences, the normalized spike components are dramatically suppressed and concurrent with enhanced hump contributions. \textbf{f}-\textbf{g}, Large-window FCS dynamics obtained with 450-nm and 400-nm CW laser irradiation.
\newpage
\clearpage

\begin{center}
	\includegraphics[width=\columnwidth]{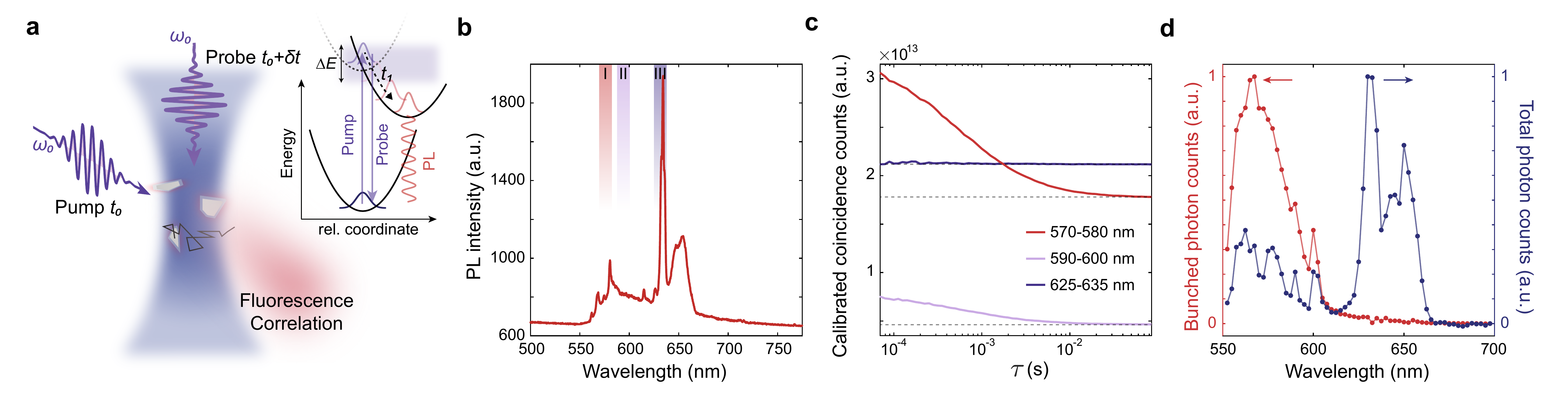}
\end{center}
\item[Fig. 3] \textbf{a}, Experimental implementation of S-SMPP. The dynamics of the photoexcited electronic state are obtained by encoding with the probe pulse in order to modify the resulting fluorescence signal. The PL background can be separated from the SQE emission using correlation detection due to its uncorrelated nature. The inset shows the energy picture of SMPP, in which the probe triggers stimulated emission of the excited state. The dashed (solid) upper curve shows the potential energy surface before (after) considering the lattice relaxation upon photoexcitation. $\Delta E$ is the excitation laser energy spread and $t_1$ marks the excited-state decay time out of $\Delta E$. \textbf{b}, PL spectrum of solution-phase hBN SQEs measured with a spectrometer. Emission from hBN SQEs is buried in the solvent Raman scattering and background light. \textbf{c}, The coincidence counts measured at different spectral ranges: 570-580 nm, 590-600 nm, and 625-635 nm marked as region I, II, and III in \textbf{b}, respectively. The trace of the region I exhibits bunching at sub-ms timescales, which reduces at the region II and disappears at the region III. The coincidence count rate is high at region III due to the ethanol and water Raman lines. \textbf{d}, The blue curve shows the total photon counts as a function of detection wavelengths, which resembles plot \textbf{b} with lowered resolution and is similarly dominated by Raman scattering lines at the region III. The red trace monitors the bunched coincidence counts at sub-ms time windows. No Raman photons are observed, and pure PL from hBN SQEs is extracted with a central wavelength of $\sim$ 570 nm.
\newpage
\clearpage

\begin{center}
	\includegraphics[width=\columnwidth]{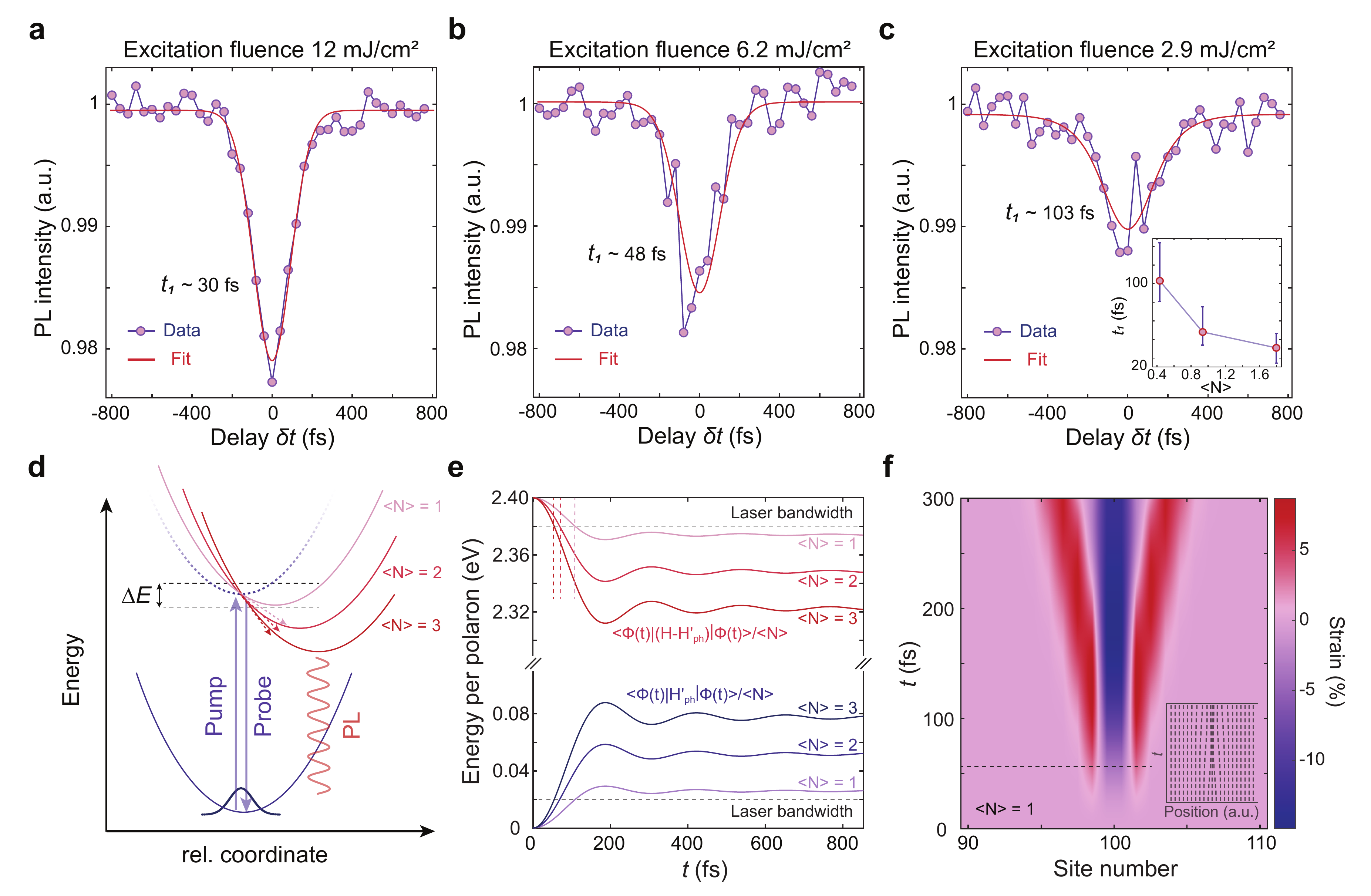}
\end{center}
\item[Fig. 4] \textbf{a}-\textbf{c}, S-SMPP measurement on hBN SQEs shows an overall PL intensity decreases near the zero pump-probe delay with a noise level of $\sim$ 0.2\%. The fit function was the convolution of the measured laser pulse intensity autocorrelation with an exponential rise function for polaron relaxation. The $t_1$ is larger at lower fluences. The inset of \textbf{c} shows the polaron relaxation time as a function of average excitation population <$N$>. The error bar is based on the fit error of the response and laser pulse duration. \textbf{d}, Schematic illustration of potential energy at increasing <$N$>, where the lower parabolic curve is the ground electronic state, the higher parabolas represent the excited states, and the red arrows show the local potential gradients. \textbf{e}, Simulated time-dependent energies of polaron formation in hBN SQEs. Normalized expectation values of the lattice Hamiltonian (less the zero point energy) $H_{ph}'$ and the remaining part of the total Hamiltonian $H-H_{ph}'$ are exhibited schematically as functions of time and varying average excitation populations. The horizontal black dashed line sketches the laser energy spread. The vertical red dashed lines marks the energy relaxation time $t_1$ out of the laser bandwidth. \textbf{f}, Simulated local strain as a function of time and spatial sites upon photoexcitation. The dashed line marks $t_1$, and the inset shows the lattice position evolution with displacements magnified by 5 times.


\newpage
\clearpage

\end{addendum}

\newpage

\section*{References}

\footnotesize

\providecommand{\noopsort}[1]{}\providecommand{\singleletter}[1]{#1}%

\newpage
\normalsize

\section*{Methods}

\noindent \textit{S-g$^{(2)}$ experimental set-up}

S-g$^{(2)}$ measurements on hBN nanoflakes in solution were completed by utilizing methods described by ref.~\cite{Beyler2014} All samples were from a prefixed solution of suspended hBN nanoflakes purchased from Graphene Supermarket. The solution was absorbed into a rectangular capillary (VitroCom, 0.100$\times$0.200 mm inner diameter) and sealed with capillary tube sealant (Fisher Scientific) to prevent solvent evaporation. Samples were excited with CW laser sources (532 nm, 450 nm, or 400 nm) using a home-built confocal epifluorescence setup based on a Olympus microscope (IX73). Excitation was focused into the sample and emission was collected using the same oil-immersion objective (Olympus, Plan Fluorite 100$\times$, NA 1.3). A 550-nm long-pass filter (Thorlabs FELH0550) and a long-pass dichroic mirror (Thorlabs DMLP550R) was used to separate excitation and emission. The excitation laser power of $\sim$ 120 (275, 245) $\mu$W at 532 (450, 400) nm was used for Fig.~2a-c (f, g). The maximal excitation power (100\%) in Fig.~2d-e was $\sim$ 240 $\mu$W. Emission was further filtered spatially using a 1:1 telescoping 200 $\mu$m pinhole and spectrally using another 550 nm long-pass filter (Thorlabs FELH0550) before being sent to a 50:50 non-polarizing beamsplitters (Thorlabs), creating two equivalent intensity beams. Each beam was focused onto a single-photon counting detector (Picoquant PDM series) using a 3 cm achromatic lens. Photon arrival times were recorded on a PicoHarp 300 (Picoquant). For correlation-based spectral measurements, multiple pairs of tunable edgepass filters (Thorlabs) were rotated for continuously varying the detection wavelengths. The setup schematic is shown in Fig.~S1. The transmission electron microscope image in Fig.~1a was performed using a transmission electron microscope (FEI Tecnai) operated at 200 keV at room temperature.\\

\noindent \textit{S-SMPP experimental set-up}

A schematic of the experimental setup is shown in Fig.~S4. Ultrafast 400-nm or 515-nm pulses were generated in a beta Barium Borate crystal by second harmonic of the output of a Ti:sapphire oscillator (KM Lab, 800 nm, 80 MHz repetition rate) or Yb amplifier (Coherent CARBIDE, 1030 nm, set to 1 MHz repetition rate). The laser pulses were split by a 50:50 beam splitter, relatively delayed by a motorized stage, and recombined by another 50:50 beam splitter. Two laser beams were spatially separated in the latter beam splitter and further cleaned up by two razor blades in both arms, which effectively removes the power changes when scanning the stages for isoenergetic SMPP. The excitation was coupled into an inverted microscope (Olympus IX73) and the excitation and emission detection geometry remains the same as S-g$^{(2)}$ setup. For S-SMPP experiment on hBN SQEs, the excitation wavelength was at 515 nm and the power was set to $\sim$ 10 pJ before the objective. For S-SMPP experiment on CdSe/CdS QD solution, the excitation wavelength was at 400 nm and the power was set to $\sim$ 5 pJ before the objective. When performing the pulse duration characterization, pump and probe pulses were intentionally overlapped on the beam splitter spatially so that the output power is modulated by the pump-probe delays.\\

\noindent \textit{Polaron formation simulation}

The polaron formation dynamics are calculated by solving the modified one-dimensional (1D) Fr\"ohlich polaron model, and the Hamiltonian of the system is given below~\cite{Brown1986,Brown1987}.
\begin{equation}
    H = \sum_{m}E_m a_m^{\dagger}a_m +\sum_{q} \hbar \omega_q (b_q^{\dagger}b_q+\frac{1}{2})+\sum_{q,m}\chi_m^q \hbar \omega_q (b_q^{\dagger}+b_{-q})a_m^{\dagger}a_m
\end{equation}
where $E_m$ is the energy at site $m$, $a_m^{\dagger}$ ($a_m$) is the exciton creation (annihilation) operator at site $m$, $b_q^{\dagger}$ ($b_q$) is the creation (annihilation) operator for the phonon mode $q$. Note that the $\sum_{q}$ refer to the summation of all $q$ within the first Brillouin zone. $\chi_m^q$ is the coupling constant between site $m$ and phonon mode $q$. In an ordered medium, $\chi_m^q=\chi^q \exp(-iqma)$. Considering only acoustic phonons, $\omega_q=\frac{2v}{a}\sin(\frac{|qa|}{2})$ and $\chi^q=\frac{2ig\sin(qa)}{\sqrt{2\hbar MN\omega_q^3}}$~\cite{Brown1987}, where $M$ is the reduced mass of boron and nitrogen atom, $N$ is the total number of cells. We estimate $N \approx 200$ using the lattice constant $a=0.25$ nm and average hBN flake size (50 nm). $v=3340$ m/s is the sound speed in hBN~\cite{Greener2018}. The lattice experiences a force from the electron with a magnitude of $g=63$ piconewtons, obtained by comparing the asymptotic phonon system energy with two times the polaron formation energy ($\approx 25$ meV) in hBN~\cite{Sio2023,Brown1986,Brown1987}. We assume that there are $N_{m'}$ excitons at site $m'$ and the excitation site to be at the center of the 1D chain, i.e., $m'=\frac{N}{2}$. Thus the system initial state is given by $\ket{\Phi(t=0)}=\frac{1}{\sqrt{N_{m'}!}}(a_{m'}^{\dagger})^{N_{m'}}\ket{0}$, where $\ket{0}$ is the common vacuum annihilated by $a_m$ and $b_q$. The time evolution of the phonon system characterizes the polaron formation time and is given by
\begin{equation}
    \begin{aligned}
        \bra{\Phi(t)} H_{ph} \ket{\Phi(t)} &=\bra{\Phi(t)} \sum_{q} \hbar \omega_q b_q^{\dagger}b_q \ket{\Phi(t)}\\
        & =\sum_{q}2\hbar \omega_q (\chi^q)^2 N_{m'}^2(1-\cos{\omega_q t})
    \end{aligned}
\end{equation}
Note that Fig.~4e plots $\frac{\bra{\Phi(t)} H_{ph} \ket{\Phi(t)}}{N_{m'}}$ to represent energy transfer to phonon subsystems contributed by each excitation. The associated lattice displacement of site $n$ is given by (details in Supplementary Note~3)
\begin{equation}
    \begin{aligned}
    d(n,t)=-\sqrt{\frac{2\hbar}{NM}} \sum_{q} \frac{\exp[-iq(n-m')a]\chi^q}{\sqrt{\omega_q}} N_{m'} (1-\cos{\omega_q t})
    \end{aligned}
\end{equation}



\subsection{Data availability.}

The data that support the findings of this study are available from the corresponding author upon request.

\newpage

\normalsize

\end{document}